# Infrared Radiation of Graphene Electrothermal Film Triggered Alpha and Theta Brainwaves

*Yanghua Lu[1, 3], Renyu Yang[1], Yue Dai[2], Deyi Yuan[1], Xutao Yu[1], Chang Liu[1], Lixuan Feng[1], Runjiang Shen[1], Can Wang[1], Shenyi Dai[4] and Shisheng Lin [1,2,3,5]\**

[1]College of Information Science and Electronic Engineering, Zhejiang University, Hangzhou, 310027, P. R. China.

[2]Hangzhou Gelanfeng Technology Co. Ltd, Hangzhou, 310051, P. R. China.

[3]Hangzhou Liangchun Technology Co. Ltd, Hangzhou, 311500, P. R. China.

[4]Hangzhou Neuro Technology Co. Ltd, Hangzhou, 310051, P. R. China.

[5]State Key Laboratory of Modern Optical Instrumentation, Zhejiang University, Hangzhou, 310027, P. R. China.

Email: shishenglin@zju.edu.cn

[*]Corresponding author.

# Abstract

The alpha and theta frequency brainwave activity in Electroencephalogram (EEG) signal has been correlated with attention, inhibitory processes, memory, perceptual abilities, and sleep. The enhanced alpha and theta brainwave activity may bring positive behavioral modifications such as promoting creativity and a quick sleep. Herein, we discover that infrared radiation from multilayer graphene electrothermal film can obviously promote the appearance of alpha and theta brainwave in human mind. In particular, the occurrence frequency of the alpha and theta waves in EEG can be effectively enhanced up to 2.3 and 3.0 times, respectively. And the duration time of the alpha and theta waves in EEG can also be effectively extended. The mechanism may be attributed to the efficient infrared radiation caused by graphene mainly focused on the range from 7 to 14 μm, coinciding with the radiation wavelength of natural human body, which can be effectively absorbed by the human skin and speed up the blood microcirculation and metabolism. The comparative effect of different working temperature and heating materials such as water, Cu and even monolayer graphene are systematically investigated, indicating the infrared radiation from the multilayer graphene electrothermal film at 50°C has the largest enhancement effect of alpha and theta brainwaves. The multilayer graphene film electrical heater represents a convenient and surprising way for triggering the alpha and theta brainwaves, which has many potential applications in the area of enlarged health cerements.

# Introduction

As a two-dimensional material, graphene has significant advantages with unique electronic, optical, thermal, and mechanical properties, such as excellent carrier mobility, strong electron-electron interaction, and ultra-high thermal conductivity.[1-6] Recently, the thermal infrared emission in carbon materials especially the graphene have been explored, which is attributed to the strong in-plane vibrational transitions of carbon atoms.[7, 8] While applying a bias voltage through the graphene film, most part of electronic energy can be transformed into infrared radiation, apart from Joule heat dissipated into the substrate contacted.[8] Infrared radiation is an invisible electromagnetic wave with a longer wavelength than visible light, which can be further subdivided into three different wavelengths: near-infrared (0.8 to 1.5 μm), middle-infrared (1.5 to 5.6 μm) and far-infrared (5.6 to 1000 μm) radiation.[9] While passing a current through the graphene, middle-infrared and far-infrared rays (4 to 20 μm) are radiated into free space.[10] Recently, we have demonstrated the fabrication of large scale multilayer graphene and patent it as the multilayer graphene film has a high far infrared emissivity over 90%,[11] which can penetrate 2 to 3 mm into human skin.[12] The infrared radiation of graphene is also matching with the human radiation wavelength (7 to 14 μm)[13], which can substantially exert strong rotational and vibrational effects at the molecular level and lead to dilation of blood vessels, enhancing the blood microcirculation and metabolism.[14-16] So the graphene infrared technique can be used in medical treatment and daily life, which can improve human health or heat preservation by enhancing blood circulation.[17, 18]

Recently, LED lights flashing at a specific 40 Hz frequency are found to significantly reduce amyloid beta plaques in the visual cortex of Alzheimer's disease mice.[19] The lamp works by inducing gamma oscillations that help the brain suppressing amyloid beta production and activate cells that destroy plaques.[20, 21] Especially, these studies could prove that Alzheimer's is not caused by the a-beta protein, but by a weakening of brainwaves.[19] Besides, the alpha and theta frequency brainwave activity has been correlated with attention, inhibitory processes, memory, perceptual abilities, and sleep.[22, 23] The decreasing power of alpha wave has been demonstrated to be the reason of aging and even the main reason of Alzheimer's disease.[24, 25] So regulating and activating the brainwaves should be an effective and potential way to treat brain diseases. The enhanced alpha and theta brainwave activity may bring positive behavioral modifications such as promoting creativity and a quick sleep.[26] Considering the positive effect of graphene on health and the effect of light on the Alzheimer's disease, we tend to ask the questions: what is the relationship between the effect of graphene infrared radiation on the human body and the activity of brainwaves?[27, 28] On the other hand, detecting and analyzing these human brain biosignals is one of the potential ways to figure out the brain operational function and human emotion[29-31], which can be implemented in such as brain injury inspection[32], relieve depression[33], creativity or cognitive functions research[34, 35]. Electroencephalogram (EEG) is a record of the oscillations of the brain electric potential[36], including detailed signals that provide information about underlying neural activities in the human brain.[37] Considering the convenience, non-invasiveness, and high temporal resolution, the analysis of fast changing dynamic EEG has become the most popular electrophysiology technique to image brain activity mapping.[38, 39]

Herein, we discover that the occurrence frequency and duration time of the alpha and theta waves in EEG can be effectively enhanced under the effect of infrared radiation from graphene electrothermal film, while the mechanism may be attributed to the blood circulation enhancing and human cell vitalizing.[40, 41] Under the bias voltage of 5.0 V, the graphene film can be heated with a controllable temperature from 40 to 60°C assisted by the feedback of NTC in the graphene film. Obvious infrared radiations with wavelengths ranging from 4 to 20 μm from the multilayer graphene film with an area of 20cm×5cm can be detected, which has a central peak spectrum locating between 7 and 14 μm. As the infrared radiation spectrum of graphene film covering the spectral range of human body at wavelengths between 7 and 14 μm, these infrared radiations from the graphene film can be effectively absorbed by the human skin, which can substantially exert strong rotational and vibrational effects at the molecular level and lead to dilation of blood vessels, enhancing the blood microcirculation and metabolism. With tightly pressing a graphene electrothermal film electrical heater imbedded in the scarf on the back of neck, the infrared radiations from the graphene film can effectively irradiate into the human skin.[42, 43] The dependence of the enhancement effect on the working temperature of graphene electrothermal film is systematically investigated. In particular, the occurrence frequency of the alpha and theta waves in EEG can be effectively enhanced up to 2.3 and 3.0 times under the optimized temperature of 50°C. And the duration time of the alpha and theta waves in EEG can also be extended. The heating effects of water, Cu and monolayer graphene are also compared with multilayer graphene film, indicating the infrared radiation from the graphene electrothermal film has the greatest enhancement effect of alpha and theta brainwaves. The controllable effect of graphene electrothermal film on the appearance of alpha and theta waves represents a convenient and non-invasive characterization tool for the regulation and activation of human brainwaves, which has potential applications such as sleep problem treatment.

## Results and Discussion

The study of the EEG is of particular interest to discover the relationship between behaviors and brain activity. In fact, the source of the EEG is the multitude of neurons and connections between them, and more precisely, it is situated on the most external layers of the brain, known as cerebral cortex. The frequencies of brain waves detected in EEG range from 0.5 to 100 Hz, and their characteristics are highly dependent on the degree of activity of the brain's cerebral cortex. Nowadays, EEG has become a powerful tool for clinical diagnosis of acute brain disorders, and EEG signals are subdivided into five different parts: Delta (0.5-4 Hz), Theta (4-8 Hz), Alpha (8-13 Hz), Beta (13-32 Hz), and Gamma (32-100 Hz) brainwaves, as shown in Figure 1a. EEG is obtained using electrodes placed on the human scalp. There are 19 general positions on the human head for brainwave test: Fp1, Fp2, F3, F4, F7, F8, Fz, T3, T4, T5, T6, C3, C4, Cz, P3, P4, Pz, O1, and O2, as shown in Figure 1b. These five brainwaves can be collected in the different positions, which represent "Asleep", "Deeply Relaxed", "Relaxed", "Focused", "Excited", respectively, as shown in Figure 1c. The general view of EEG's signal processing mode and flow diagram is shown in Figure 1d. The human brain signals collected with electrodes were transformed to digital signals, and then filtered with balance electrodes signals to eliminate noise signals, outputting effective EEG signals under the effect of the graphene film with effective infrared radiation.

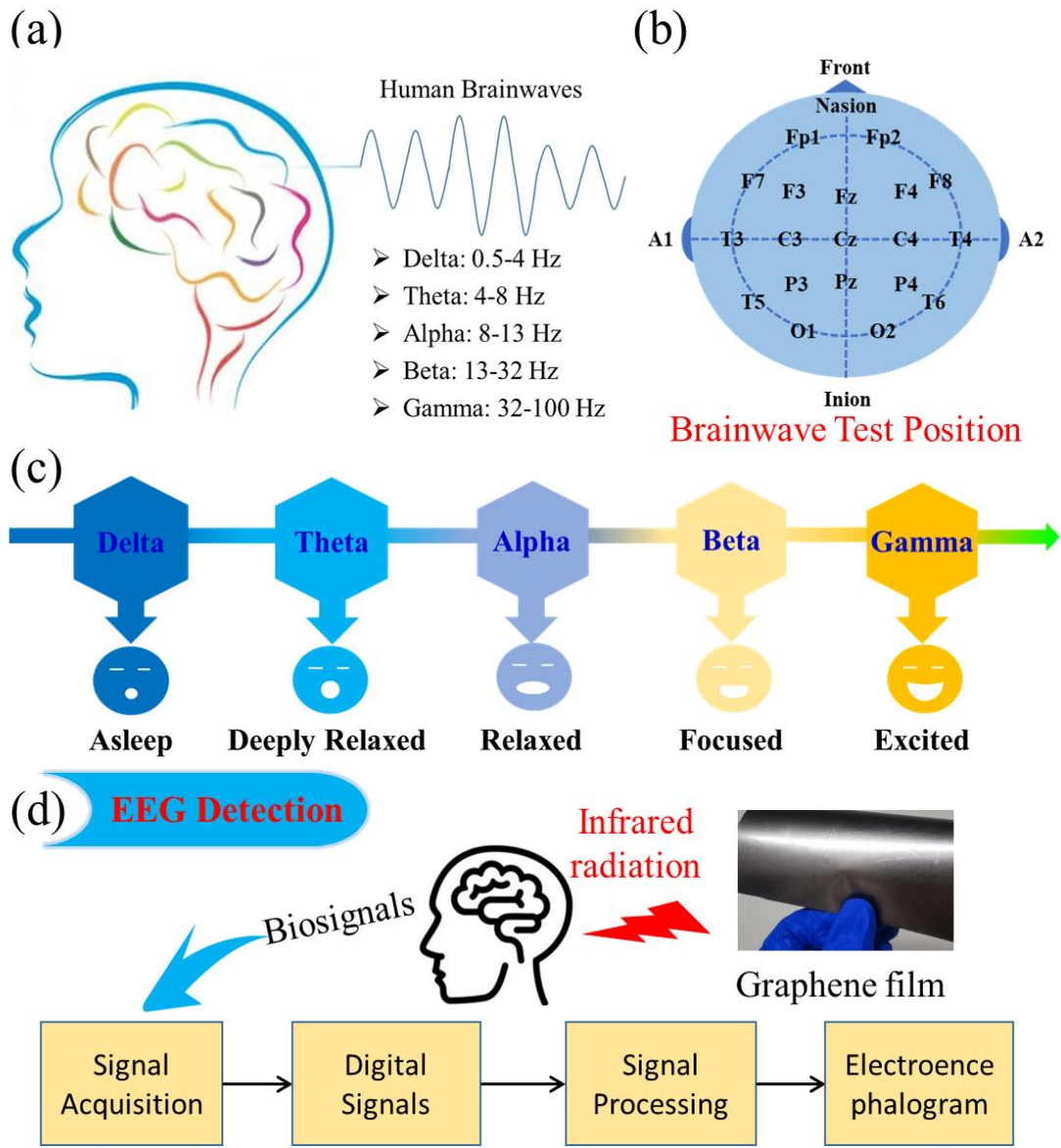

**Figure 1. Schematic diagrams of EEG detection of human brainwaves.** (a) Human brainwaves and the frequency of five type brainwaves. (b) Brainwaves test position. (c) Different emotion states of five type brainwaves. (d) General view of EEG's signal processing mode and flow diagram of signal detection.

During the measurement of EEG signals, a graphene electrothermal film electrical heater was tightly pressed to the back of the neck, which is imbedded in the scarf. As shown in Figure 2a, grid-shaped graphene electrothermal film with the area of 20cm×5cm was used. The graphene film fabricated by the tape casting method has excellent flexibility and mechanical stability, which indicates its superiority for wearable heating device.[44, 45] The top and cross-section view of Scanning electron microscopy (SEM) image of the graphene film is shown in Figure 2b and Figure 2c, respectively. Figure 2c reveals that the graphene film was composed of vertically stack graphene flake layers. The Raman spectrum of the graphene film at 25°C was also measured and shown in Figure 2d, which has three peaks named as D-peak (1348 cm$^{-1}$), G-peak (1584 cm$^{-1}$), and 2D-peak (2724 cm$^{-1}$), as shown in Figure 2d. In particular, the weak D-peak showed the high quality and limited crystal defects of the graphene film. Under the bias voltage of 5V, a temperature of 40°C to 60°C can be achieved under the feedback of NTC in the graphene film, as shown in Figure 2e. Broad infrared radiation can be detected at wavelengths between 4 and 20 μm[10, 11], which can penetrate 2 to 3 mm into human skin through the scalp[12], as shown in Figure 2f. Especially, the peak energy of the infrared radiation spectrum of graphene film covers the spectral range of human body at wavelengths between 7 and 14 μm as shown in the shaded area of Figure. 2c, so the infrared radiation from the graphene film can be effectively absorbed by the human skin, which can substantially lead to dilation of blood vessels, enhancing the blood microcirculation and metabolism. As shown in Figure S1 and S2, a graphene electrothermal film of 50.7°C embedded in a scarf with a temperature of 41.8°C was tightly pressed to the back of the neck, and the temperature of human skin was lifted from 37.5°C to 40.2°C, attributing to the infrared radiations from graphene film.

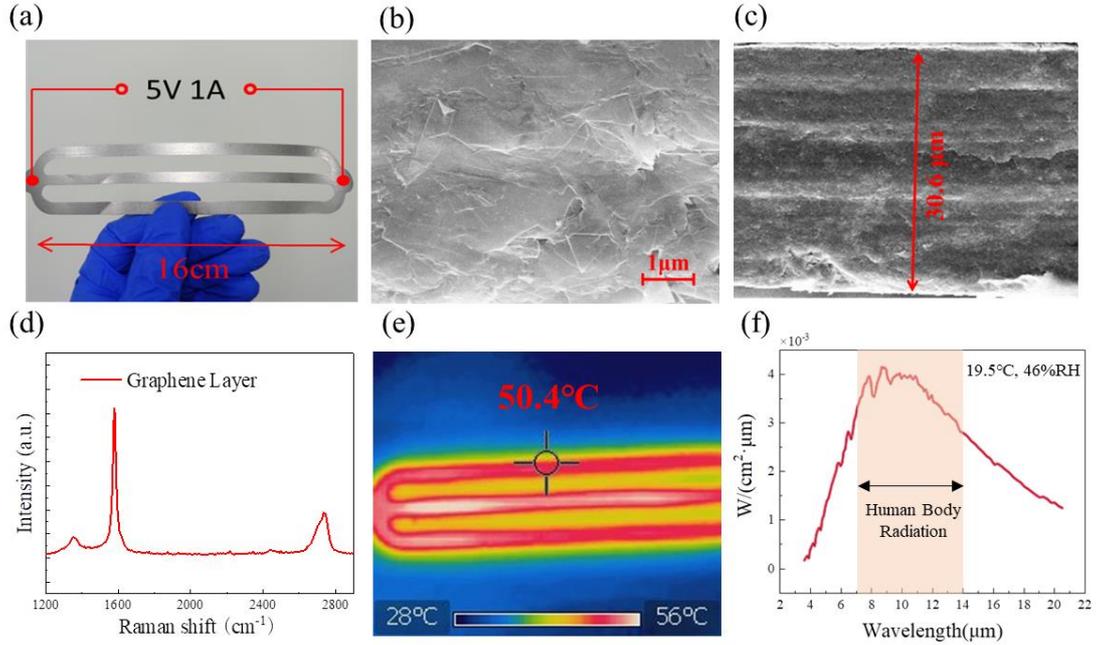

**Figure 2. Optical and electrical properties of the graphene electrothermal film.** (a) Optical picture of the grid-shaped graphene electrothermal film. SEM images of the graphene film in the (b) horizontal section and (c) vertical section. (d) Raman spectrum of the graphene film at a room temperature of 25°C. (e) Infrared image of the graphene film under a bias voltage of 5.0 V. (f) Infrared radiation spectrum of the graphene electrothermal film.

Alpha brainwaves are considered to be the brainwave state of relax and calmness, which can be measured with EEG.[46, 47] For the alpha wave test, point O1 and point O2 were chosen to record the electric signal, and the GND and REF worked as a contrast zero potential or reference potential. The "O" of O1 and O2 means occipital lobe, an area of human brain associated with vision. According to the brainwaves test position in Figure 1b, we found these two points place on the occipital bone at the back of the brain. The measurements are carried out a perfectly quiet and dark room. During the measurement, the participants always closed eyes, which avoided the interruption of the visual system and reduced the noise information generated by visual processing in the occipital lobe. With tightly pressing a graphene electrothermal film electrical heater imbedded in the scarf on the back of neck, the infrared radiation from the graphene film can effectively irradiate into the human skin. The alpha wave of EEG with duration of 60 seconds was measured before and after the heating of graphene electrothermal film for 3 minutes, as shown in Figure 3a and 3b, respectively. We discovered that the occurrence frequency of the alpha waves in EEG effectively increased from 3.0 to 7.0 times/min after participants are exposed to the infrared radiation from graphene electrothermal film. And the typical alpha wave before and after graphene heating is shown in Figure 3c and 3d, respectively.

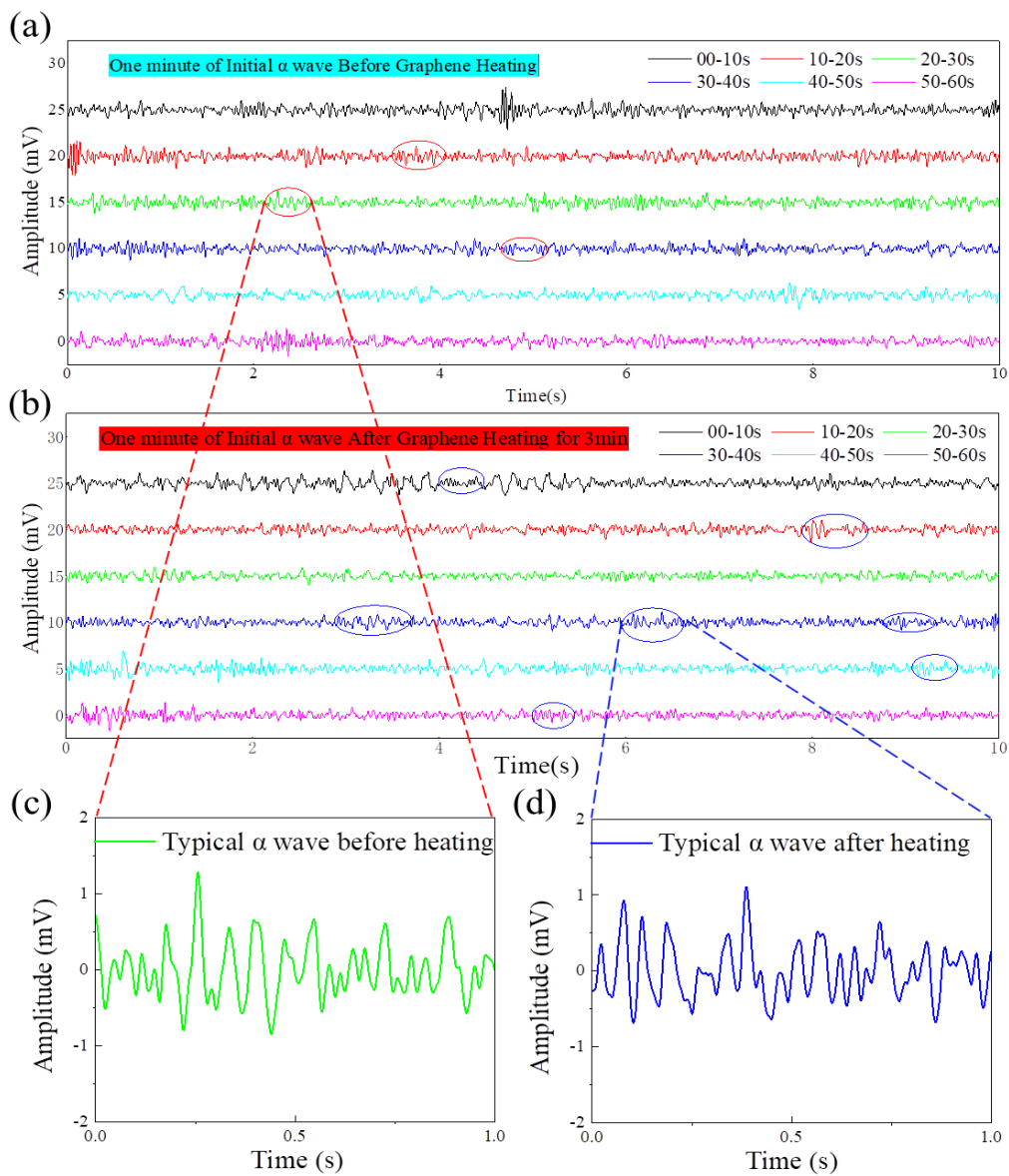

**Figure 3. Alpha waves in EEG signals.** (a) One minute of initial alpha waves from O1 or O2 positions before graphene heating. (b) One minute of alpha waves from O1 or O2 positions after the heating effect of graphene electrothermal film for 3min. (c) Typical alpha wave before graphene heating. (d) Typical alpha wave after heating.

The rationality of the existence of alpha brainwaves is one of the basic states of the brain, which is innate in our human brain but easily hampered by the stresses of environment and life, leading to stress and anxiety diseases. The alpha waves in EEG signals indicate less anxiety and stress, leading to higher immunity and creativity. We mainly referred to two items in data analysis: the occurrence frequency and the active time of the alpha waves before and after the heating of graphene electrothermal film. As shown in Figures 4a, the occurrence frequency of alpha waves in EEG signals for one minute before and after heating of graphene electrothermal film were measured as 3.0 and 7.0 times/min, respectively. It can be found that the mean time, maximum time, and total time of alpha waves in EEG signals were effectively increased from 0.5/0.5/1.4 s to 0.6/0.8/4.0 s, as shown in Figure 4b.

Furthermore, the dependence of the enhancement effect on the working temperature of graphene electrothermal film was systematically investigated. As shown in Figure 4c, the occurrence frequency of alpha waves under the temperature of 25/40/45/50/55°C was 3.0/4.0/5.0/7.0/5.0 times/min, respectively. The occurrence frequency increased and then decreased with the increase of temperature, and 50°C was found as the optimal choice, in which the human brain can achieve the most comfortable state. To eliminate the environmental influence and prove the infrared radiation plays a key role in the enhancement effect of alpha brainwaves, we have carried out experiments on the relationship between the occurrence frequency and different heating material, such as water, Cu film and even monolayer graphene film.

As shown in Figure 4d, the occurrence frequency of alpha brainwaves under the heating of water, Cu, monolayer graphene and multilayer graphene at 50°C were 3.0, 4.0, 5.0 and 7.0 times/min, respectively. To control variables in the test, the resistance of the Cu, monolayer graphene and multilayer graphene used was 5.0 Ω. The detailed signal analysis of alpha waves in EEG signals under the heating of water Cu and monolayer graphene electrothermal film were shown in Figure S3. Compared with multilayer graphene electrothermal film used, the water, Cu and monolayer graphene electrothermal film behaved limited enhancement effect, indicating the most obvious role of infrared radiation from the multilayer graphene electrothermal film in the enhancement effect of alpha brainwaves. The typical alpha waves in EEG signals before and after the heating of Cu, monolayer graphene or multilayer graphene electrothermal film were shown in Figure S4. Power spectral density of alpha wave in EEG signals before and after heating of graphene electrothermal film was also compared. As shown in Figure S5, with the heating of graphene electrothermal film, the power spectral density of alpha wave was enhanced and the peak frequency shifts to lower part. Especially, the occurrence of alpha wave during the frequency of 8-11 Hz indicates the human mind was blankly confused and consciousness, which was gradually moving towards the deeply relaxed.[48] Alpha waves are the "frequency bridge" between conscious mind (Beta waves) and unconscious mind (Theta waves). So the heating of graphene electrothermal film can help to calm down and promote a deeper sense of relaxation and contentment, which induces more alpha waves and even theta waves.

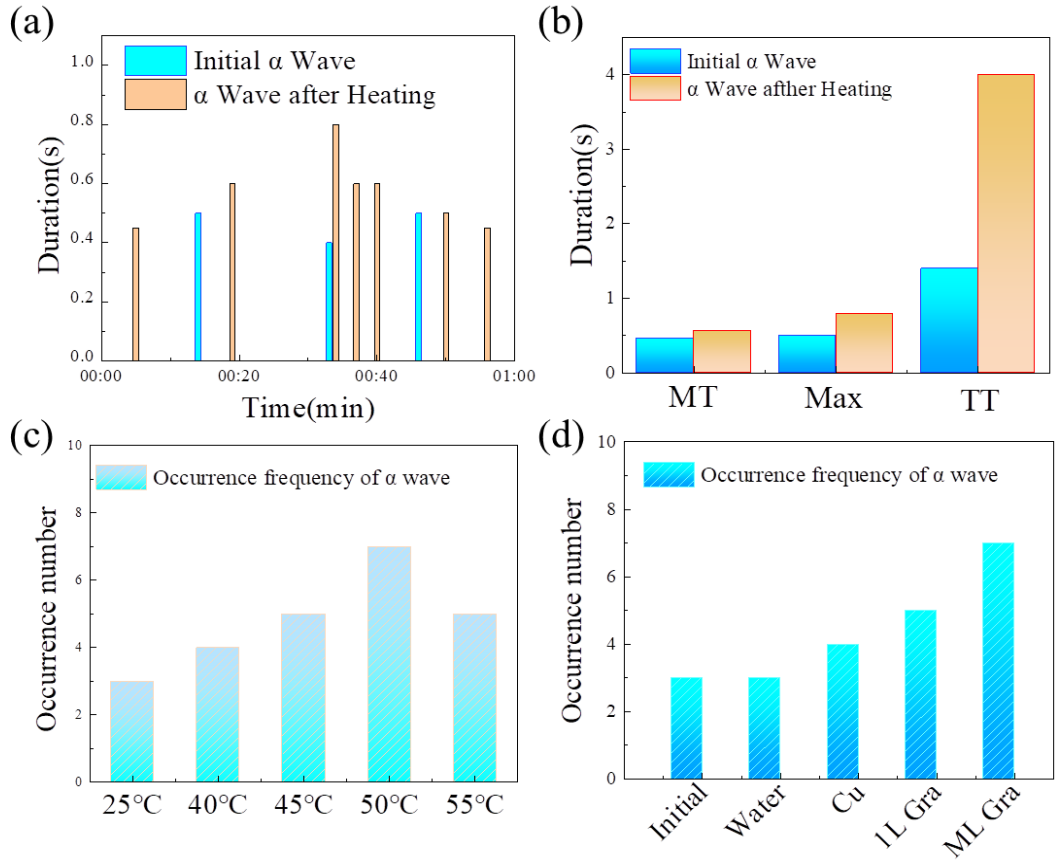

**Figure 4. Signal analysis of alpha waves in EEG signals.** (a) Occurrence frequency of alpha wave in EEG signals for one minute before and after heating of graphene electrothermal film. (b) Occurrence time contrast of alpha wave in EEG signals before and after heating of graphene electrothermal film. (c) Occurrence frequency of alpha wave in EEG signals under different heating temperature. (d) Occurrence frequency of alpha wave in EEG signals under the heating of water, Cu, monolayer graphene and multilayer graphene electrothermal film.

Theta waves usually occur when deeply relaxed or near the time fall sleeping, which are commonly found in trance or hypnosis. In addition, theta waves are important for triggering deep memories and strengthening long-term memories. The theta waves of the brain make it easier for patients to receive hypnosis, which have been widely used in sleep or mental health related treatments. Therefore, the theta waves of EEG were also measured before and after the heating of graphene electrothermal film. As alpha and theta waves are always found in different areas, point F3 and point F4 were chosen to record the electric signal for the theta wave test, and point Cz worked as the reference potential. For the convenience of the theta wave EEG signals detection, the participants always closed the eyes and breathed regularly. To assist in the observation of the theta waves, the participants were required to hold deep breath and keep the breath frequency in 13-15 times per minute. As the theta waves are infrequent in adult, the closed eyes and deep breathe can reduce the effect of environment and enhance the occurrence of theta waves. As shown in Figure 5a and 5b, the occurrence frequency of the theta waves in EEG was effectively enhanced from 2.0 to 6.0 times/min, under the effect of infrared radiation from graphene electrothermal film. And the typical theta waves before and after graphene heating are shown in Figure 5c and 5d, respectively.

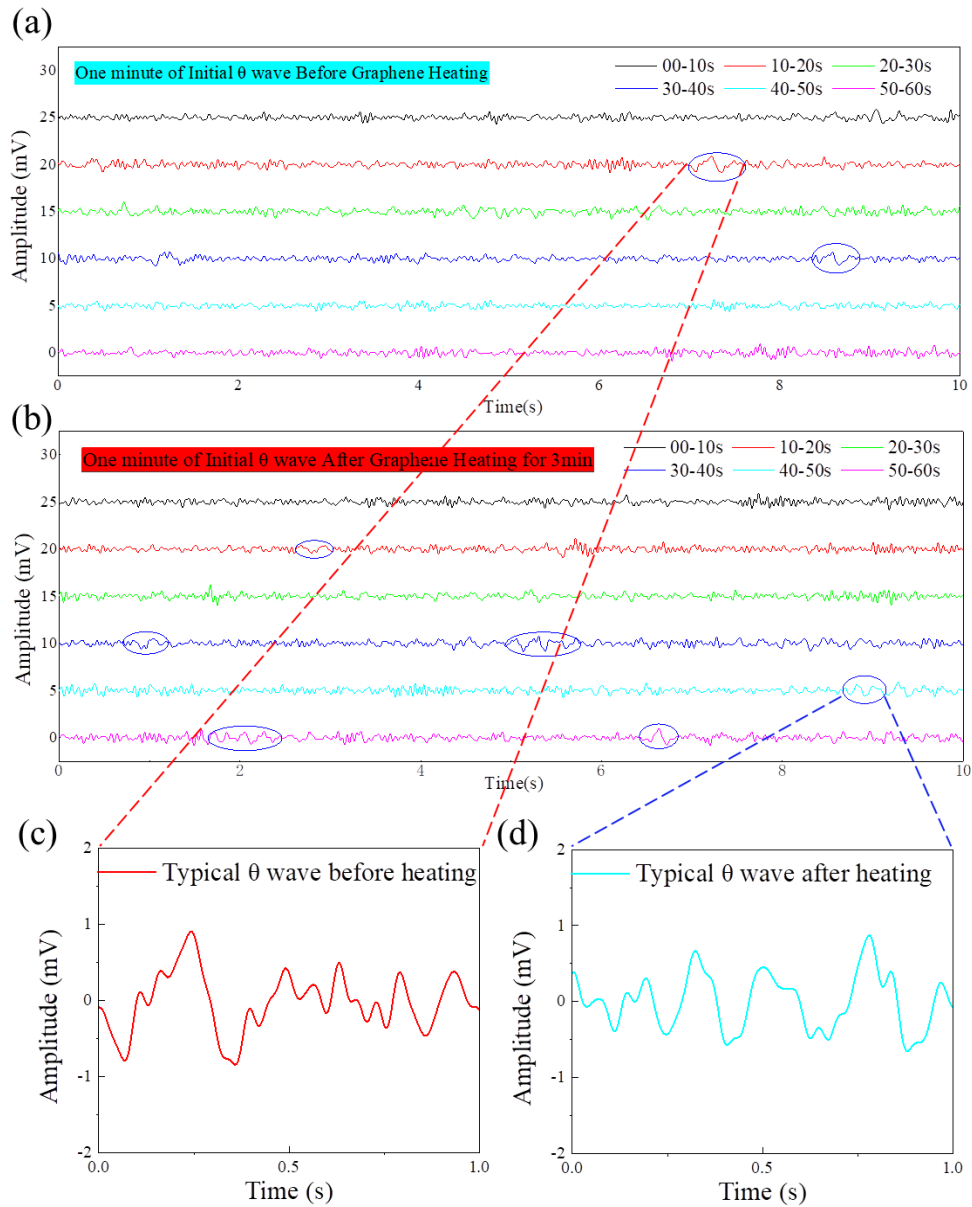

**Figure 5. Theta waves in EEG signals.** (a) One minute of initial theta wave from F3 or F4 positions before graphene heating. (b) One minute of theta wave from F3 or F4 positions after the heating of graphene film for 3 min. (c) Typical theta wave before graphene heating. (d) Typical theta wave after graphene heating.

The heating of graphene electrothermal film can help to promote a deeper sense of relaxation and fall asleep, inducing more theta waves. To reveal the enhancement effect of graphene film on the theta waves, we mainly referred to two items in data analysis: the occurrence frequency and the lasting time of the theta waves before and after the heating of graphene electrothermal film. The occurrence frequency of theta wave in EEG signals for one minute before and after heating of graphene electrothermal film was shown in Figures 6a, which is 2.0 and 6.0 times/min, respectively. It can be found that the mean time, maximum time, and total time of theta waves in EEG signals were be effectively enhanced from 0.4/0.4/0.8 s to 0.6/0.8/3.3 s, as shown in Figure 6b. Furthermore, the dependence of the theta waves enhancement effect on the working temperature of graphene electrothermal film was systematically investigated. As shown in Figure 6c, the occurrence frequency of theta waves under the temperature of 25/40/45/50/55°C was 2.0/3.0/5.0/6.0/3.0 times/min, respectively. The occurrence frequency increased and then decreased with the increase of temperature, and 50°C was found as the optimal choice, in which the human brain achieved the most comfortable state for sleeping. For comparison, we also have carried out experiments on the relationship between the occurrence frequency and different heating material, such as water Cu film, and monolayer graphene film. As shown in Figure 6d, the occurrence frequency of theta brainwaves under the heating of water, Cu, monolayer graphene and multilayer graphene at 50°C were 2.0, 3.0, 4.0 and 6.0 times/min, respectively. To control variables in the test, the resistance of the Cu, monolayer graphene and multilayer graphene used was 5.0 Ω. The detailed signal analysis of theta waves in EEG signals under the heating of water Cu and monolayer graphene electrothermal film were shown in Figure S6. Compared with multilayer graphene electrothermal film, the water, Cu and monolayer graphene electrothermal film behaved limited enhancement effect, indicating the key role of infrared radiation from the multilayer graphene electrothermal film in the enhancement effect of theta brainwaves. The typical theta waves in EEG signals before and after the heating of Cu, monolayer graphene or multilayer graphene electrothermal film were shown in Figure S7. Power spectral density of theta wave in EEG signals before and after heating of graphene electrothermal film was also compared. As shown in Figure S8, with the heating of graphene electrothermal film, the power spectral density of theta wave was enhanced and the peak frequency decreases. Especially, the occurrence of theta wave during the frequency of 4-6 Hz indicates the human mind was deeply relaxed and hypnagogic.[48] Theta waves can improve the creativity, wholeness, and intuition, which can also involve in the restorative sleep. Therefore, the graphene electrothermal film electrical heater represents a convenient and non-invasive characterization tool for the regulation and activation of human brain biosignals, which has potential applications in human brain activity promotion and sleep problem treatment.

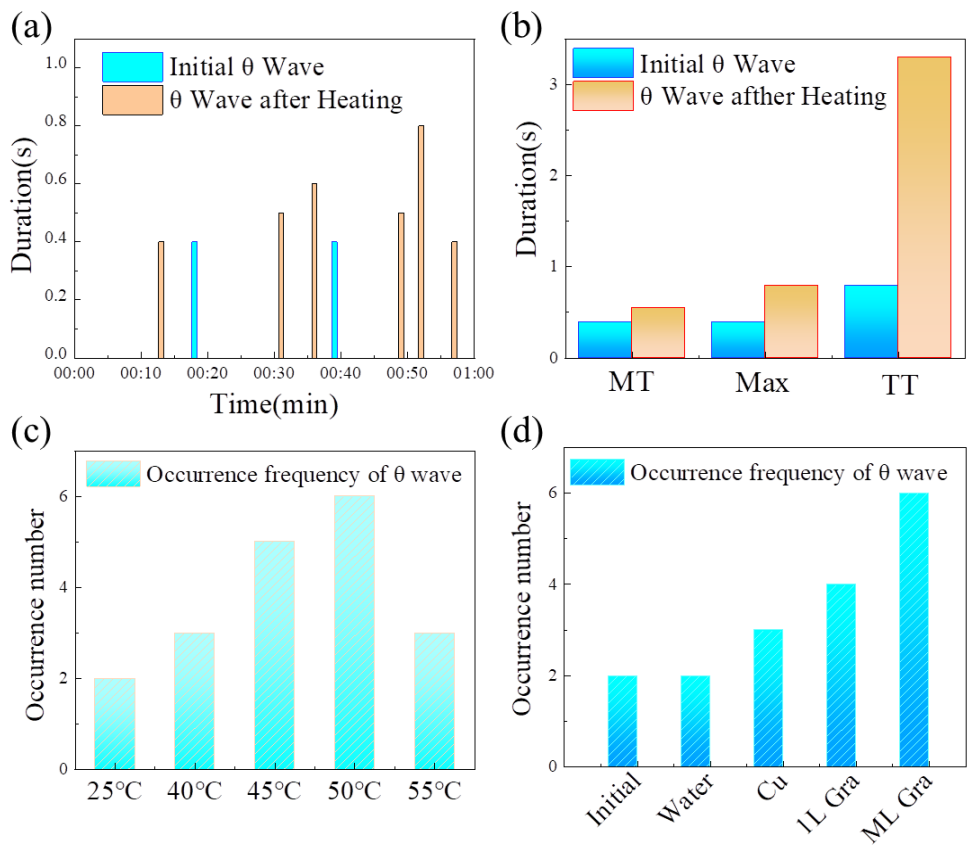

**Figure 6. Signal analysis of theta wave in EEG signals.** (a) Occurrence frequency of theta wave in EEG signals for one minute, before and after heating of graphene

electrothermal film. (b) Occurrence time contrast of theta wave in EEG signals before and after heating of graphene electrothermal film. (c) Occurrence frequency of theta wave in EEG signals under different heating temperature. (d) Occurrence frequency of theta wave in EEG signals under the heating of water, Cu, monolayer graphene and multilayer graphene electrothermal film.

## Conclusion

In summary, we discover that the occurrence frequency and duration time of the alpha and theta waves in human brain can be effectively enhanced with tightly pressing a graphene electrothermal film electrical heater imbedded in the scarf on the back of neck. We have uniquely point out, the occurrence frequency of the alpha and theta waves in EEG can be effectively increased up to 2.3 and 3.0 times under the optimized temperature of 50°C. And the duration time of the alpha and theta waves in EEG can also be extended effectively. The heating effects of water, Cu and even monolayer graphene are also compared with multilayer graphene film, indicating the key role of infrared radiation from the graphene electrothermal film in the enhancement effect of alpha and theta brainwaves. The graphene electrothermal film electrical heater represents a convenient and non-invasive characterization tool for enhancing the neurocognitive function associated with systems such as memory and attention and also the detection of EEG, which has many potential applications in the area of enlarged health cerements.

## Experimental Section

***Device fabrication and measurement:*** Before the test, the participants' heads were treated with alcohol wipes to remove excessive sebum and keratin. The Au disk electrodes with diameter of 10mm were placed on the matched points. All electrodes were fixed on the scalp with medical tape and conductive paste. We got the digital signals from iNeuro system and dealt with raw data by using MATLAB (2019a, MathWorks, USA). For alpha wave test, theparticipants' foreheads were treated with alcohol wipes, the same as point O1 and point O2. The points we chose were REF, GND, O1&O2. REF and GND were on either side of the forehead, which were acting as a contrast zero potential or reference potential. During the test, the participants should close their eyes and keep awake. Every single test lasted about 10 minutes, of which the last 8 minutes were the data of using the infrared radiation. The alpha and theta waves are always found in different areas. In the theta wave test, we chose Cz as reference point and F3&F4 as test points. The data were dealt with in the same method as the alpha wave test, while the test process had a bit of difference. To assist the observation of the theta rhythm, the participants were required to hold deep breath in the test, and keep the breath frequency in 13-15 times per minute.

*Characterization analysis:* The iNeuro system includes two main software, iNeuro.Client.Collect.exe(V3.1.0) and iNeuro.Client.Replay.exe(V3.0.3). The former is used for EEG collecting, and another could replay the stored EEG data. The operating system could set data collecting parameter and do some general processing, like changing the data flow, setting EEG leads and filter. We got the raw data as digital signals from iNeuro system. We selected the channels we used in test and then imported the data into MATLAB. After the signal filtering processing, effective EEG signals can be achieved. The waveforms we got were bandpass filtered at 3-20 Hz by MATLAB's "filter" function. The EEG power spectral density (PSD) is a frequency-domain parameter, which directly reflects the values of different frequencies in the frequency domain. We used P-welch function in MATLAB to quantify the brainwave data after preprocessing, and got the PSD graphs of alpha rhythm and theta wave from each EEG dataset. As for the parameters in P-welch function, we set 10-s hanning window，5120 points of Nfft，512-hz sampling frequency and 50% overlap.

## Supporting Information

Supporting Information is available online.

## Author Contributions

S. Lin designed the experiments, participated the experiments, analyzed the data, conceived the study, and wrote the paper. Y. Lu, R. Yang, Y. Dai assisted to design and carry out the experiments, discuss the results. D. Yuan, X. Yu, H. Zheng, Y. Yan, C. Liu, Z. Yang, L. Feng, R. Shen and C. Wang discussed the results and assisted with experiments. All authors contributed to the preparation of the manuscript.

## Acknowledgements

S. S. Lin thanks the support from the National Natural Science Foundation of China (Grant Nos. 51202216, 51551203 and 61774135), Special Foundation of Young Professor of Zhejiang University (Grant No. 2013QNA5007), and Outstanding Youth Fund of Zhejiang Natural Science Foundation of China (Grant No. LR21F040001). Y. H. Lu thanks the support from the China Postdoctoral Science Foundation (Grant No. 2021M692767).

## Conflict of Interest

The authors declare no conflict of interest.